# Giant strain gradient elasticity in SrTiO$_3$ membranes: bending versus stretching

V. Harbola*[‡,1,2], S. Crossley*[‡,2,3], S. S. Hong[2,3], D. Lu[1,2], Y. A. Birkhölzer[4], Y. Hikita[2] & H. Y. Hwang*[2,3]

[1]Department of Physics, Stanford University, Stanford, California 94305, USA
[2]Stanford Institute for Materials and Energy Sciences, SLAC National Accelerator Laboratory, 2575 Sand Hill Road, Menlo Park, California 94025, USA
[3]Department of Applied Physics, Stanford University, Stanford, California 94305, USA
[4]Department of Inorganic Materials Science, Faculty of Science and Technology and MESA+ Institute for Nanotechnology, University of Twente, P.O. Box 217, 7500 AE Enschede, The Netherlands

[‡]These authors contributed equally: V. Harbola and S. Crossley
Email: varunh@stanford.edu, samuelcrossley@gmail.com, hyhwang@stanford.edu

*Informal summary:* We show how crystalline nano-membranes of flexoelectric SrTiO$_3$ mechanically behave like this familiar toy, in that they are much easier to stretch than bend.

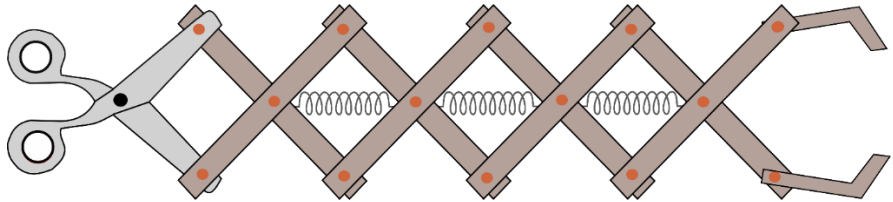

**Young's modulus determines the mechanical loads required to elastically stretch a material, and also, the loads required to bend it, given that bending stretches one surface while compressing the opposite one. Flexoelectric materials have the additional property of becoming electrically polarized when bent[1,2]. While numerous studies have characterized this flexoelectric coupling[3–6], its impact on the mechanical response, due to the energy cost of polarization upon bending[7], is largely unexplored. This intriguing contribution of strain gradient elasticity[8,9] is expected to become visible at small length scales where strain gradients are geometrically enhanced, especially in high permittivity insulators[1,2,7]. Here we present nano-mechanical measurements of freely-suspended SrTiO$_3$ membrane drumheads. We observe a striking non-monotonic thickness dependence of Young's modulus upon small deflections. Furthermore, the modulus inferred from a predominantly bending deformation is three times larger than that of a predominantly stretching**

**deformation for membranes thinner than 20 nm. In this regime we extract a giant strain gradient elastic coupling of ~2.2 µN, which could be used in new operational regimes of nano-electro-mechanics.**

The last two decades have seen tremendous advances in materials synthesis and engineering, allowing for precise control in creating and manipulating nano-mechanical structures of various crystalline materials[10–14]. These capabilities have enabled nano-elastic studies with geometries ranging from rods to cantilevers to drumheads. It is remarkable how classical linear elastic theory, founded on Hooke's law, can be readily extended to the analysis of the mechanical behavior of such nanostructures. Large monotonic size effects have been observed in Young's modulus $E$ for materials such as silicon[10], silicon nitride[11], and others (see Supplementary Information and Extended Data Fig. 1), which generally scale as $\Delta E \propto \pm 1/t$ with respect to the bulk limit, where $t$ is the thickness. These size effects are attributed to surface elasticity arising from a difference in the local $E$ between the surface and interior regions, where the positive (negative) sign implies a stiffer (less stiff) surface.

A natural extension beyond linear response is to consider that, in addition to the strain $\varepsilon$, the strain gradient $\nabla \varepsilon$ may contribute to the elastic energy[8,9,15,16] (Fig. 1a, b). Namely, the Gibbs free energy density $\phi$ can be expressed as

$$\phi = \frac{1}{2}E\varepsilon^2 + \frac{1}{2}K(\nabla\varepsilon)^2 \qquad (1)$$

where $K$ characterizes the energy coupling strength to the strain gradient, and $\frac{1}{2}K(\nabla\varepsilon)^2$ defines strain gradient elasticity (SGE). (For simplicity we suppress the full tensor form in this schematic equation.) In general SGE is expected to be very small, of order $K \sim 1$ nN[16], and virtually undetectable in the elastic response of bulk crystals. However, because strain gradients induced by bending scale inversely with size[1,2,8], the contribution of SGE should vary (exclusively positive) as $\propto +1/t^2$ under bending (Fig. 1a, b). But even with this geometric scaling, SGE has been deemed to likely be inconsequential on the nanoscale.

So where might one look for large SGE? Possible microscopic origins are an active field of research[9,16], with a number of different mechanisms for SGE proposed. Among them, flexoelectricity is considered promising. Flexoelectricity is the electric polarization $P$ that develops in an insulator when bent, where the flexoelectric coefficient $\mu$ characterizes the coupling strength of $P$ to the strain gradient, which breaks inversion symmetry[1,2]. Flexoelectricity is therefore permitted for all crystal structures, although it is typically only noticeable in insulators with large lattice contributions to the dielectric response[3,4]. This connection with permittivity $\chi$ is apparent from the relation $\mu = \chi f$, where $f$ is the flexocoupling coefficient. Thus it is natural that complex oxides with large $\chi$ such as $SrTiO_3$ (with $\chi \sim 300\epsilon_0$ and $f \sim 2.6$ V at room temperature, where $\epsilon_0$ is the vacuum permittivity) are among the best studied bulk flexoelectrics[3,5].

Flexoelectricity is one pathway to SGE because the induced polarization adds an electrostatic energy $P^2/2\chi$ that must also be overcome while bending the crystal[6,7]; and it is notable that simple uniform stretching does not have this contribution. Since $P = \mu \nabla \varepsilon$, the electrostatic energy term is precisely of the form of SGE, with $K = \mu^2/\chi$. These considerations (the materials, the geometry) suggest that nanoscale flexoelectrics could exhibit significant contributions to elasticity arising from SGE[7,9]. Unfortunately, measuring the mechanical response of thin films does not access this possibility, as they are bound on the substrate on which they are grown, diminishing the strain gradient by the substrate thickness. However, the recent development of freestanding crystalline oxide membranes with nanoscale thickness[17] provides new experimental access to this nano-mechanical regime, which we study here.

We fabricated freely suspended crystalline SrTiO$_3$ membranes of variable thickness ($4 \leq t \leq 98$ nm) via pulsed laser deposition and subsequent lift-off using a water-soluble epitaxial buffer layer (see Methods). These membranes were then transferred onto 200 nm thick Si$_3$N$_4$ membranes with arrays of holes of radius $R$ (0.25, 1, or 5 μm), thereby creating circular drumheads (Fig. 1c). Transmission electron microscopy has shown that these drumheads remain highly crystalline down to 2 nm[18]. For measurement of elastic properties, we used atomic force microscopy (AFM) techniques (Fig. 1d) developed for two-dimensional (2D) materials such as graphene[13,19], MoS$_2$ (ref.[14,20]), and h-BN[21]. Fine and direct control of drumhead deformation within a regime of repeatable and reversible elasticity measurements, and encompassing both bending and stretching, mark this as an ideal platform for exploring SGE in these SrTiO$_3$ membranes.

A total of 168 SrTiO$_3$ drumheads were measured (Extended Data Table 1). For each drumhead, the AFM tip was used to probe a 32×32 lateral 2D array of positions encompassing the freely suspended drumhead and the surrounding Si$_3$N$_4$-supported region (Fig. 1c). At each point a force-position $F(d)$ experiment was performed for both loading and unloading of the AFM tip ($F$ is the force deflecting the AFM tip, measured through $d_{tip}$, at sample vertical position $d$). The $F(d)$ data were then converted into $F(\delta)$ data ($\delta$ is the deflection of the SrTiO$_3$ membrane from its neutral position, directly under the tip) by accounting for the mechanical compliances of the Si$_3$N$_4$ and the AFM tip itself (see Methods). Averaging over loading and unloading $F(\delta)$ curves yielded a single-valued monotonic function from which the elastic properties of the drumheads were deduced.

Representative traces from these studies are shown in Fig. 2a, b. For simplicity, we focus our discussion on center-point loading. The $F(\delta)$ data are linear for thicker drumheads, becoming nonlinear for thinner drumheads and larger deflections. These trends are similar to those predicted[22] and observed[13,14] previously in 2D materials. Since symmetry considerations forbid even-power terms in the polynomial expansion of $F(\delta)$ under center-point loading about $\delta = 0$, to leading order the nonlinear elastic drumhead response is given by[13,14,19,22]

$$F(\delta) = \left[\frac{4\pi E_{lin} t^3}{3(1-\nu^2)R^2} + \pi T\right]\delta + \left[\frac{cE_{cub}t}{R^2}\right]\delta^3 \qquad (2)$$

where $T$ is the membrane pre-tension, $\nu$ is the Poisson ratio, and $c$ is a numerical coefficient for a given experimental geometry (see Supplementary Information). $E_{lin}$ and $E_{cub}$ correspond to effective Young's modulus for the deformations giving rise to linear and nonlinear $F(\delta)$, respectively. In the absence of SGE (or other higher order contributions to elasticity[8,23]), $E_{lin}$ and $E_{cub}$ are both equal to $E$, and both can be used to measure its value. However, as we will demonstrate, for thin SrTiO$_3$ membranes $E_{lin}$ and $E_{cub}$ are quite distinct, and vary significantly with membrane thickness $t$.

A three-dimensional finite element simulation (3D FES) (see Methods) readily captures both the linear and nonlinear regimes, and gives physical insight into their origins. Whereas the linear $F(\delta)$ is due to a bending deformation, the nonlinear $F(\delta)$ corresponds to an increasing degree of stretching (Fig. 2c), all within the elastic response. In the perturbative limit of purely linear $F(\delta)$, the mechanical response is determined by $T$ and $E_{lin}$, which can be identified as the intrinsic bending rigidity of the material[19,24]. By going beyond center-point loading, we can independently determine both $T$ and $E_{lin}$ solely from the linear response (allowing for a separable measurement of $E_{cub}$). Therefore, these quantities are extracted by fitting the full spatially-resolved 2D compliance ($\frac{d\delta}{dF}\big|_{F=0}$) dataset to the numerical solution of the biharmonic equation for bending (Fig. 2d) (see Methods). Furthermore, the results of our 3D FES provide a 'lookup table' for the coefficient $c$, enabling us to relate the experimentally determined cubic contribution to $F(\delta)$ for center-point loading, to $E_{cub}$ (see Methods).

Using this analysis, we extract the thickness dependence of the effective moduli corresponding to the linear (bending dominated) and cubic (stretching dominated) $F(\delta)$ response (Fig. 3). We find that $E_{lin}$ is strongly non-monotonic in $t$, showing a minimum value of ~50 GPa at $t = 31$ nm. Above this thickness $E_{lin}$ increases asymptotically towards the bulk value (~300 GPa)[25]. Below this thickness, $E_{lin}$ increases sharply to ~250 GPa at $t = 14$ nm, the smallest drumhead thickness where the purely linear regime of $F(\delta)$ could be independently probed. However, the same 14 nm thick drumheads show $E_{cub} \sim E_{lin}/3$, exemplifying a robust trend of $E_{cub} < E_{lin}$ for thicknesses below the minimum in $E_{lin}$ ($t = 31$ nm). The non-monotonicity and the existence of a minimum in the thickness dependence of $E_{lin}$ can be understood as a crossover from a strain gradient dominated deformation response, to a surface elasticity dominated response, as the thickness is increased. These results reveal that the SrTiO$_3$ membranes in this thin regime are indeed intrinsically stiffer on bending as compared to stretching, and their mechanical response cannot be accounted for by simple linear elasticity. We also found that $E_{cub}$ decreases with increasing thickness up to $t = 24$ nm, the largest drumhead thickness where nonlinearity in $F(\delta)$ was experimentally accessible. For samples with $t < 14$ nm, the linear regime of $F(\delta)$ was not accessible for evaluation of $E_{lin}$, as the region of apparent linearity only presented at forces ~1 nN, which is below the snap-in force on the cantilever (Fig. 2a, b and Extended Data Fig. 3).

To the extent of our knowledge, no other material has been shown to exhibit this highly non-monotonic thickness dependence in the linear elastic response. However, the distinctive size-scaling of surface elasticity versus SGE ($\pm 1/t$ and $+1/t^2$ respectively) provides a straightforward explanation: negative surface elasticity dominates for $t > 31$ nm, and SGE for $t < 31$ nm (Fig. 3). The presence of SGE also explains the observation that $E_{cub} < E_{lin}$ for 14 nm $\leq t \leq 24$ nm, given that SGE only enhances the intrinsic bending rigidity $E_{lin}$. It should be noted that $E_{cub}$ provides only an upper bound on effective Young's modulus for collinear stretching (see Supplementary Information), since the nonlinear response includes both stretching and bending deformations (Fig. 2c). Thus bending also contributes to the thickness variation in $E_{cub}$, albeit to a lesser extent than $E_{lin}$. By considering the additional bending rigidity under the assumption of a uniform SGE, we deduce a value of $K = 2.2 \pm 0.7$ μN for $t = 14$ nm, which is 3 orders of magnitude higher than those tabulated theoretically for typical materials[16].

Returning to the connection between SGE and flexoelectricity, we can relate the SGE coefficient $K$ to the flexoelectric coefficient $\mu$ by $K = \mu^2/\chi$ (see Methods). This corresponds to effective coefficients of $|\mu| \sim 86 \pm 13$ nCm$^{-1}$ and $|f| \sim 32 \pm 5$ V. It should be noted that other effects, such as surface piezoelectricity, can also contribute to SGE with the same thickness scaling as flexoelectricity, but with different microscopic origin[1]. We reiterate that these parameters are derived presuming that the observed SGE arises solely from a uniform flexoelectric response. Taken at face value, the flexocoupling strength is very large given that the intrinsic lattice contributions to $f$ are expected to be bounded by Kogan's phenomenological estimate of 1-10 V[26,27]. Our results add to a growing body of evidence that this limit can be surpassed, particularly in high-permittivity ceramic materials and at the nanoscale[4,28]. In particular, it is worth noting that the values we derive here using purely elasticity measurements corroborate enhancements measured electrically.

A deeper microscopic understanding of nanoscale flexoelectric enhancement, as well as other potential contributions to SGE (see Supplementary Information), clearly warrants further investigation. Nevertheless, these results visibly demonstrate that nanoscale $SrTiO_3$ exhibits sharply different bending and stretching rigidity, raising the possibility for new dynamics in nano-electro-mechanical devices[29]. Furthermore, we note that even for materials with much smaller intrinsic flexoelectric response, its dominant scaling in the 2D limit suggests it may be qualitatively relevant for the physics of the 2D crumpling transition[30,31]. Finally, the rapid development of synthetic techniques for preparing membranes of many complex oxides[32,33] and their heterostructures[17,34], including magnetic[34,35] and multiferroic[36] materials, provides the opportunity for flexo-coupling to degrees of freedom beyond electric polarization.

**References**

1. Zubko, P., Catalan, G. & Tagantsev, A. K. Flexoelectric effect in solids. *Annu. Rev. Mater. Res.* **43**, 387–421 (2013).

2. Yudin, P. V. & Tagantsev, A. K. Fundamentals of flexoelectricity in solids. *Nanotechnology* **24**, 432001 (2013).

3. Zubko, P., Catalan, G., Buckley, A., Welche, P. R. L. & Scott, J. F. Strain-gradient-induced polarization in $SrTiO_3$ single Crystals. *Phys. Rev. Lett.* **99**, 167601 (2007).

4. Ma, W. & Cross, L. E. Flexoelectric polarization of barium strontium titanate in the paraelectric state. *Appl. Phys. Lett.* **81**, 3440–3442 (2002).

5. Bhaskar, U. K. *et al.* A flexoelectric microelectromechanical system on silicon. *Nat.*

*Nanotechnol.* **11**, 263–266 (2016).

6. Cordero-Edwards, K., Domingo, N., Abdollahi, A., Sort, J. & Catalan, G. Ferroelectrics as smart mechanical materials. *Adv. Mater.* **29**, 1702210 (2017).

7. Majdoub, M. S., Sharma, P. & Cagin, T. Enhanced size-dependent piezoelectricity and elasticity in nanostructures due to the flexoelectric effect. *Phys. Rev. B* **77**, 125424 (2008).

8. Lam, D. C. C., Yang, F., Chong, A. C. M., Wang, J. & Tong, P. Experiments and theory in strain gradient elasticity. *J. Mech. Phys. Solids* **51**, 1477–1508 (2003).

9. Stengel, M. Unified ab initio formulation of flexoelectricity and strain-gradient elasticity. *Phys. Rev. B* **93**, 245107 (2016).

10. Li, X., Ono, T., Wang, Y. & Esashi, M. Ultrathin single-crystalline-silicon cantilever resonators: Fabrication technology and significant specimen size effect on Young's modulus. *Appl. Phys. Lett.* **83**, 3081–3083 (2003).

11. Babaei Gavan, K., Westra, H. J. R., Van Der Drift, E. W. J. M., Venstra, W. J. & Van Der Zant, H. S. J. Size-dependent effective Young's modulus of silicon nitride cantilevers. *Appl. Phys. Lett.* **94**, 233108 (2009).

12. Chen, C. Q., Shi, Y., Zhang, Y. S., Zhu, J. & Yan, Y. J. Size dependence of Young's modulus in ZnO nanowires. *Phys. Rev. Lett.* **96**, 075505 (2006).

13. Lee, C., Wei, X., Kysar, J. W. & Hone, J. Measurement of the elastic properties and intrinsic strength of monolayer graphene. *Science* **321**, 385–388 (2008).

14. Castellanos-Gomez, A. *et al.* Elastic properties of freely suspended $MoS_2$ nanosheets. *Adv. Mater.* **24**, 772–775 (2012).

15. Mindlin, R. D. Micro-structure in linear elasticity. *Arch. Ration. Mech. Anal.* **16**, 51–78 (1964).

16. Maraganti, R. & Sharma, P. A novel atomistic approach to determine strain-gradient elasticity constants: Tabulation and comparison for various metals, semiconductors, silica, polymers and the (Ir) relevance for nanotechnologies. *J. Mech. Phys. Solids* **55**, 1823–1852 (2007).

17. Lu, D. *et al.* Synthesis of freestanding single-crystal perovskite films and heterostructures by etching of sacrificial water-soluble layers. *Nat. Mater.* **15**, 1255–1260 (2016).

18. Hong, S. S. *et al.* Two-dimensional limit of crystalline order in perovskite membrane films. *Sci. Adv.* **3**, eaao5173 (2017).

19. Poot, M. & Van Der Zant, H. S. J. Nanomechanical properties of few-layer graphene membranes. *Appl. Phys. Lett.* **92**, 63111 (2008).

20. Bertolazzi, S., Brivio, J. & Kis, A. Stretching and breaking of ultrathin $MoS_2$. *ACS Nano* **5**, 9703–9709 (2011).

21. Li, C., Bando, Y., Zhi, C., Huang, Y. & Golberg, D. Thickness-dependent bending modulus of hexagonal boron nitride nanosheets. *Nanotechnology* **20**, 385707 (2009).

22. Komaragiri, U., Begley, M. R. & Simmonds, J. G. The mechanical response of freestanding circular elastic films under point and pressure loads. *J. Appl. Mech.* **72**, 203–212 (2005).

23. Eringen, A. C. On differential equations of nonlocal elasticity and solutions of screw dislocation and surface waves. *J. Appl. Phys.* **54**, 4703–4710 (1983).

24. Landau, L. D., Pitaevskii, L. P., Kosevich, A. M. & Lifshitz, E. M. *Theory of Elasticity*. (Butterworth-Heinemann, 2012).

25. Carpenter, M. A. Elastic anomalies accompanying phase transitions in $(Ca,Sr)TiO_3$ perovskites: Part I. Landau theory and a calibration for $SrTiO_3$. *Am. Mineral.* **92**, 309–327

(2007).

26. Kogan, S. M. Piezoelectric effect during inhomogeneous deformation and acoustic scattering of carriers in crystals. *Sov. Phys. Solid State* **5**, 2069–2070 (1964).

27. Tagantsev, A. K. Piezoelectricity and flexoelectricity in crystalline dielectrics. *Phys. Rev. B* **34**, 5883–5889 (1986).

28. Das, S. *et al.* Enhanced flexoelectricity at reduced dimensions revealed by mechanically tunable quantum tunnelling. *Nat. Commun.* **10**, 537 (2019).

29. Davidovikj, D. *et al.* Ultrathin complex oxide nanomechanical resonators. Preprint at https://arxiv.org/abs/1905.00056 (2019).

30. Deng, S. & Berry, V. Wrinkled, rippled and crumpled graphene: An overview of formation mechanism, electronic properties, and applications. *Mater. Today* **19**, 197–212 (2016).

31. Nelson, D. R. & Peliti, L. Fluctuations in membranes with crystalline and hexatic order. *J. Phys. Paris* **48**, 1085–1092 (1987).

32. Dong, G. *et al.* Super-elastic ferroelectric single-crystal membrane with continuous electric dipole rotation. *Science* **366**, 475–479 (2019).

33. Bakaul, S. R. *et al.* Single crystal functional oxides on silicon. *Nat. Commun.* **7**, 10547 (2016).

34. Kum, H. S. *et al.* Heterogeneous integration of single-crystalline complex-oxide membranes. *Nature* **578**, 75–81 (2020).

35. Paskiewicz, D. M., Sichel-Tissot, R., Karapetrova, E., Stan, L. & Fong, D. D. Single-crystalline SrRuO$_3$ nanomembranes: A platform for flexible oxide electronics. *Nano Lett.* **16**, 534–542 (2016).

36. Ji, D. *et al.* Freestanding crystalline oxide perovskites down to the monolayer limit. *Nature* **570**, 87–90 (2019).

**Methods**

**Fabrication of SrTiO₃ drumheads**

The fabrication of freestanding SrTiO$_3$ (Extended Data Fig. 2) was originally described in ref.[17], and subsequently refined for obtaining large-area crack-free membranes[18] as utilized here.

*Pulsed laser deposition*: Growth substrates were (001)-oriented 5 mm × 5 mm × 0.5 mm single crystal polished and annealed SrTiO$_3$ (Shinkosha Ltd, Japan). A lamp heater was used to heat the substrate to ~900 °C in ~5×10$^{-6}$ Torr of pure O$_2$ gas. The substrate was annealed under these conditions for ~45 minutes. A pulsed excimer laser (wavelength of 248 nm and pulse repetition rate < 2 Hz) was imaged on the surface of ceramic targets located 5 cm below and opposite to the polished surface of the heated substrate, creating an ablation plume. Epitaxial bilayer films of 8 nm thick Sr$_3$Al$_2$O$_6$ and *t* nm thick SrTiO$_3$ were grown on the substrate by ablating targets of Sr$_3$Al$_2$O$_6$ and SrTiO$_3$ at a laser fluence of ~1.25 Jcm$^{-2}$ and ~0.45 Jcm$^{-2}$ respectively.

*Release and transfer of SrTiO$_3$ freestanding films*: A ~200 nm thick polymer support layer of polymethyl methacrylate (PMMA) was deposited atop the as-grown bilayer film of Sr$_3$Al$_2$O$_6$ and SrTiO$_3$ by spin coating and subsequent baking at ~150 °C. The entire structure of growth substrate, bilayer oxide film and polymer support were then immersed in deionized water at room temperature. The water dissolved away the sacrificial 8 nm thick Sr$_3$Al$_2$O$_6$ layer, allowing the growth substrate to be physically removed, leaving only the *t* nm thick SrTiO$_3$ film attached to the polymer support. This structure was transferred onto commercial 200 nm thick Si$_3$N$_4$ membranes with an array of circular holes with diameter 500 nm, 2 µm and 10 µm (Norcada NH005D05, NH050D2, NH050D1032 respectively). The polymer support was removed by washing in acetone followed by isopropanol. Finally, oxygen plasma ashing was used for three minutes to remove polymer residue. This yielded arrays of freestanding drumheads of SrTiO$_3$ supported by a much larger and less mechanically compliant membrane of Si$_3$N$_4$, itself supported by a bulk Si frame.

*Characterization of as-grown SrTiO$_3$ films*: During film growth, the film surface and thickness was characterized by reflection high-energy electron diffraction (RHEED). Oscillations in the intensity of the RHEED reflection verified that a layer-by-layer film growth mode was maintained for all samples. The thickness of the films was examined by atomic force microscopy (AFM) on the transferred film and agreed with the intended growth thickness.

*Characterization of transferred SrTiO$_3$ freestanding films*: AFM in tapping mode was used to map the topography of the transferred SrTiO$_3$ freestanding films (Extended Data Fig. 2c). AFM in the vicinity of a cleaved edge of the transferred film permitted film thickness to be evaluated (Extended Data Fig. 2c). The single-crystal silicon support structure of the Si$_3$N$_4$ membrane was (001)-oriented and provided an alignment feature for X-ray diffraction (XRD) of the transferred SrTiO$_3$ membrane (Extended Data Fig. 2d). No parasitic phases or orientations were apparent in the diffraction data. We also compare XRD of the as-grown and transferred SrTiO$_3$ film (Extended Data Fig. 2d) with the SrTiO$_3$ substrate dominating the signal from the as-grown sample.

**Force microscopy measurement**

Force microscopy was conducted using an AFM (Asylum Research Cypher) using carbon-coated tips (BudgetSensors Multi75DLC) with a tip radius of ~15 nm, force constant of $k \sim 3$ Nm$^{-1}$, and free-space resonance frequency of ~75 kHz. For elastic measurements, the tip radius does not have a notable effect on the measurement as long as it is much smaller than the drumhead radius[13]. During force spectroscopy measurement, the AFM vertical drive-axis impacted the sample surface by moving a distance $d$ into the stationary tip (Fig. 1d), causing the tip to deflect a distance $d_{tip}$. For each drumhead, the AFM tip indented at 32×32 lateral positions encompassing the freely suspended region and the surrounding Si$_3$N$_4$-supported region. This deflection was sensed via the vertical difference signal of the optical position-sensitive detector of the AFM (Extended Data Fig. 3). A 'snap-in' feature in the data (Extended Data Fig. 3a) enabled identification of the moment of contact of the tip to the sample surface, upon loading and unloading. In addition to the 'snap-in' hysteresis, a small hysteresis due to tip bowing and friction was also notable between the loading and unloading force-position $F(d)$ curves. Except for the purpose of identifying the neutral position of SrTiO$_3$ drumheads (see below), averaging over loading and unloading $F(d)$ curves was employed to obtain a single-valued monotonic function[37]. The maximum force applied on freestanding membranes ranged from 20-200 nN depending on drumhead thickness and radius. Repeatability of measurements confirmed the absence of plastic deformation or slippage of the membrane on the nitride surface. We note that 24 nm thick membranes were the thickest we could push into the nonlinear $F(\delta)$ regime, with a maximum force of 200 nN. Increasing the applied force also increases the hysteresis between approach and retract traces due to higher cantilever bowing forces[37]; thus we limited the maximum force applied to 200 nN.

To calibrate $d_{tip}$, the tip was impacted into the rigid surface of the Si support structure,

such that $d_{\text{tip}} = d$ was enforced (Extended Data Fig. 3). Force constant $k$ was determined for each cantilever by the thermal tuning procedure[38]. This enabled the applied force on the cantilever $F$ to be determined in all subsequent experiments, via the relation $F = kd_{\text{tip}}$. Each indentation on the 32x32 positions thus yielded an $F(d)$ dataset for both loading and unloading. We note that the tip force response was linear in $d_{\text{tip}}$ up to the largest force of 200 nN used in this study.

On impacting the tip into compliant surfaces, namely the freely suspended $SrTiO_3$ drumheads, and $Si_3N_4$-supported regions (Extended Data Fig. 3c), the sample surface deflection $\delta_{\text{sur}}$ was evaluated as $|d-d_{\text{tip}}|$ (Extended Data Fig. 3d). When impacting the tip onto the $SrTiO_3$ drumhead, the supporting $Si_3N_4$ membrane was also deflected, though by a much smaller amount for the membranes studied (up to $t$ = 98 nm). (Above this value, the increasing contribution from the finite thickness (200 nm) and stiffness of the $Si_3N_4$ membrane limits the reliability of our measurement.) Subtraction of the $Si_3N_4$ component yielded $F(\delta)$, where $\delta$ is the under-tip deflection of the $SrTiO_3$ drumhead alone (Extended Data Fig. 3e).

For analysis of linear $F(\delta)$, the zero position for $\delta$ could be arbitrarily specified. However, in order to quantitatively extract nonlinearity in $F(\delta)$, it is necessary to identify the neutral position of the $SrTiO_3$ drumhead in order to accurately fit a cubic function. We identified this neutral position with the inflection point that was observed, shown in the unloading curve (Extended Data Fig. 3f). It should be noted that from the raw data, as in Extended Data Fig. 3c, it is not readily apparent where the actual neutral position of the membrane is, and to what tip deflection it would correspond. $d_{\text{tip}} = 0$ is the neutral point of the cantilever but when the cantilever makes contact with the membrane, in general that may not be the actual neutral point of the membrane. Therefore, by fitting the raw data to $F - F_0 = a(\delta' - \delta_0) + b(\delta' - \delta_0)^3$ where $\delta'$ is the x-coordinate of the raw data, we accounted for this inherent constant offset both in position and force, allowing us to identify the inflection point as $(F_0, \delta_0)$. The true deflection $\delta$ of the membrane is then given by $\delta' - \delta_0$. The axes in Extended Data Fig. 3e, f have been shifted post analysis so that the origin coincides with the membrane neutral position. We reiterate that the non-linearity in $F(\delta)$ response arises due to a geometric effect. As the thickness of the membrane is increased, the bending rigidity increases as $t^3$, resulting in a significantly smaller $\delta$ for the same force. The coefficient of the cubic term only varies linearly with thickness (equation (2) in the main text), so as thickness increases, the onset of visible nonlinearity in the $F(\delta)$ response emerges at a much larger $\delta$.

**Three-dimensional finite element simulation**

A three-dimensional finite element simulation (3D FES) was used to deduce the behavior of an ideal SrTiO$_3$ drumhead within a framework of isotropic, linear elasticity. While SGE was not explicitly incorporated, our 3D FES achieved two important purposes: a qualitative understanding that the nonlinear $F(\delta)$ is the result of a transition to stretching-dominated deformations, and a quantitative determination of the numerical coefficient $c$ for our specific drumheads, which was used to interpret our experimental data. Mesh generation was automated, permitting drumhead geometry, material properties, drumhead pre-tension and loading to be systematically varied. The finite element solver (Calculix[39]) employed a nonlinear iterative approach to simulate drumhead deformation under center-point loading.

The finite element mesh employed 5760 20-node brick elements (Extended Data Fig. 4a). The mesh consisted of a uniformly elastic circular plate with a freestanding region of radius $R$ (grey-shaded volume in Extended Data Fig. 4a) and a Si$_3$N$_4$-supported region (green-shaded volume in Extended Data Fig. 4a) which extended a distance $0.5R$ beyond the boundary of the freestanding region. The effect of the $Si_3N_4$ support was simulated by imposing clamped boundary conditions on the red-shaded face indicated in Extended Data Fig. 4b. The finite element simulation imposed a force $F$ onto the red node indicated in Extended Data Fig. 4b, representing the force of the AFM tip on the membrane. The response of all nodal displacements *u* in the simulation were captured (Extended Data Fig. 4c). The displacement of the blue node illustrated in Extended Data Fig. 4b defined the simulated under-tip drumhead displacement $\delta$, i.e. the quantity to be compared directly with the experimentally measured $\delta$ under center-point loading (Extended Data Fig. 4d). It may be seen from Extended Data Fig. 4d that the finite element model readily captured both linear and nonlinear $F(\delta)$ behavior observed in experiments.

To analyze the nature of the drumhead deformations induced by the applied force, we considered the vector displacements *u* of specific nodes at the upper and lower surfaces along a particular radial line (Extended Data Fig. 5a). Each set of four adjacent nodes defined a rectangular 'cell' of undeformed corner coordinate vectors $\boldsymbol{x}_i = (x_i, z_i)$ and deformed corner coordinates $\boldsymbol{x}_i' = (x_i', z_i')$ [$0 \leq i \leq 3$] (Extended Data Fig. 5b), where $\boldsymbol{u}_i = \boldsymbol{x}_i' - \boldsymbol{x}_i$. Thus the strains along the radial x-direction of the upper and lower surfaces of each cell were given by $\varepsilon_u = (|\boldsymbol{x}_1' - \boldsymbol{x}_0'| - |\boldsymbol{x}_1 - \boldsymbol{x}_0|)/|\boldsymbol{x}_1 - \boldsymbol{x}_0|$ and $\varepsilon_l = (|\boldsymbol{x}_3' - \boldsymbol{x}_2'| - |\boldsymbol{x}_3 - \boldsymbol{x}_2|)/|\boldsymbol{x}_3 - \boldsymbol{x}_2|$, respectively (strains along the out-of-plane *z*-direction were negligibly small in comparison to those along the *x*-direction, thus we refer to the strain in the *x*-direction as simply 'the strain'). We assumed that internal cell strain $\varepsilon(z)$ linearly interpolated between $\varepsilon_l$ and $\varepsilon_u$, and we thus calculated $\langle \varepsilon^2 \rangle$, the average squared strain present in the cell, through integration of this linear interpolant and subsequent normalization by cell height *t*. Similarly, we calculated the average squared strain gradient along the *z*-direction $\langle (\nabla \varepsilon)^2 \rangle = |\varepsilon_l - \varepsilon_u|^2/t^2$. (Due to the high aspect ratio of our drumheads, strain gradients with respect to the *x*-direction were negligible in comparison with those with respect to the *z*-direction, thus here and elsewhere we refer to the derivative of strain along the *x*-direction with respect to the *z*-direction as simply 'the strain gradient').

The cells whose corners were defined by the nodes indicated in Extended Data Fig. 5a are represented in Extended Data Fig. 5c. For each of these cells, $\langle\varepsilon^2\rangle$ and $\langle(\nabla\varepsilon)^2\rangle$ are plotted in Extended Data Fig. 5d. In a pure bending deformation, the ratio of these two quantities is fixed, whereas in a pure stretching deformation, $\langle(\nabla\varepsilon)^2\rangle$ is zero throughout. Thus, a plot of their ratio (Extended Data Fig. 5e) represents a proxy for the degree of bending relative to the degree of stretching present along the radial profile of the simulated $SrTiO_3$ drumhead. It may be seen from Extended Data Fig. 4d that bending is the dominant deformation at the center and outer perimeter of the drumhead, whereas stretching is dominant in intermediate areas.

The ratio $\langle(\nabla\varepsilon)^2\rangle/\langle\varepsilon^2\rangle$ of Extended Data Fig. 5e are replotted as a color scale to a plot of drumhead deflection *u*(*r*), in Extended Data Fig. 5f, for three different regimes of applied force *F*, corresponding to the transition from linear to nonlinear *F*(*δ*). In the linear regime (*F* = 2 nN)*,* the deformation is entirely dominated by bending. Thus, our simulation results show that the linear to nonlinear transition in *F*(*δ*) corresponds to a crossover from a pure bending deformation towards a hybrid deformation in which stretching plays an increasing role. These deformations correspond to what we experimentally determine to present distinct effective Young's moduli $E_{lin}$ and $E_{cub}$, respectively. We also note that the peak strain observed near the tip was 0.3% for these simulations.

**Numerical solution to the biharmonic equation**

We consider perturbative bending deformations of a uniform plate of thickness *t*,

Young's modulus $E$ and Poisson ratio $v$ and thus bending rigidity $D = Et^3/[12(1-v^2)]$ about its neutral plane. Mechanical deformation $u(r)$ under loading pressure $p(r)$ (where $r$ describes the in-plane position) are described by the solution to the general biharmonic equation[24]:

$$D\nabla^4 u(r) - T\nabla^2 u(r) = p(r) \qquad (3)$$

where $T$ is the uniform pre-tension of the plate. It is convenient to express this as two coupled Poisson equations:

$$D\nabla^2 u - Tu = f \qquad (4)$$

$$\nabla^2 f = p(r) \qquad (5)$$

where $f$ is an auxiliary variable. We used a finite element approach to solve the coupled equations numerically for a circularly clamped plate of radius $R$, by imposing boundary conditions $u = 0$ and $\partial u/\partial r = 0$ at $r = R$. The circular membrane was represented by a 2D mesh of ~4000 symmetric three-node triangular elements (doubling or halving the number of elements affected the calculation time but did not significantly change the results). The function $p(r)$ describing the pressure profile applied by the AFM tip was modeled as a Gaussian function centered at tip position $r_0$, with a width 1/50th that of the drumhead radius (20 nm for $R = 1$ μm), comparable to the quoted radius of the tip (~15 nm) used in experiments. The modeled tip radius changed to 100 nm in the modeling for 5 μm holes. However, we compared these solutions to solutions with higher mesh densities and a 25 nm tip and the resulting deformations were the same, albeit the latter taking longer to compute. The force from the tip at these relative scales of tip radius and drumhead size essentially acts like a point force. The integral of $p(r)$ over all area was equal to the applied force $F$.

The above mesh, and specified values of $T$, $E$, $p$ and $r_0$, were passed to a finite element solver (FEniCS[40]) to yield simulated $u(r)$ for those $T$, $E$, $p$ and $r_0$ (Extended Data Fig. 6a, b). The finite element calculation was repeated at 15 equally spaced values of $r_0$ between zero and $R$. For the $u(r)$ calculated for the 15 different values of $r_0$, we evaluated 15 distinct values of $u(r_0)$, i.e. the under-tip deflection probed experimentally as $\delta$. Given that the bending (equation (5)) is only valid in the linear perturbative regime of $F(\delta)$, we normalize $\delta$ by $F$ to yield the local compliance as a function of $r$ (Extended Data Fig. 6c). These simulated $d\delta/dF(r)$ correspond to particular values of $T$ and $E$ provided as parameters to the finite element solver. We used an iterative least squares optimization to fit the simulated $d\delta/dF(r)$ to the experimental $d\delta/dF(r)$, in order to extract the $T$ and $E$ implied by the experimental data (the simulation input parameters $T$ and $E$ were varied at each iteration, recalculating the simulated $d\delta/dF(r)$ each time). Extended Data Fig. 6d illustrates how the functional forms of the normalized simulated $d\delta/dF$ were modified by increasing $T$. These simulated data were highly comparable to simulated data of other groups for 2D materials[41], and to our 3D FES in the perturbative linear regime.

**'Lookup table' for nonlinear response**

Previous studies[22] suggest a closed form for $c$ from equation (2) of the main text, for very large deformations, which has been used widely in the 2D materials community[13,14,20]. However, we are typically in the crossover regime between linear and cubic $F(\delta)$, and no analytical form for $c$ exists in this crossover[20,42] (Fig. 2a, b). Thus, we used our 3D FES to investigate the variation of $c$, determining a 'lookup table' for $c$ as a function of the fixed parameter $t$. This approach is valid given that we show that $c$ is a weak function of the additional parameters $R$, $T$ and particularly $E$ itself (for our 3D FES based on isotropic elasticity). Within the parameter space ($R$, $E$, $T$, $t$) we kept three parameters constant, and varied the fourth while determining $c$ from simulation, as illustrated in Extended Data Fig. 7a-d. The maximum force used to simulate was 200 nN which bounds the experimental force range within which we operate.

From the simulation results shown in Extended Data Fig. 7a-c, we determined that variation of $c$ was within ~12% on varying ($R$, $E$, $T$). We therefore write:

$$c(R, E, T, t) \approx c(t) \qquad (6)$$

where $c(t)$ is a coefficient determined empirically by a least square fit to the finite element simulation $F(\delta)$ by a cubic polynomial (Extended Data Fig. 7d). The simulated $c(t)$ data shown in Extended Data Fig. 7d constituted a 'lookup table' enabling the experimentally determined coefficient of the $\delta^3$ term to be mapped to an effective $E_{cub}$ through equation (2) of the main text. The ~12% variation shown in Extended Data Fig. 7a-c is incorporated into the error bar that was used in quoted values of $E_{cub}$ (Fig. 3). The $c(t)$ used for each thickness is calculated using the experimental value of $R$. It is notable that at low values of $t$ the coefficient $c(t)$ determined by our finite element simulation (Extended Data Fig. 7d) extrapolated well to the value of ~1 derived by others[22] for the atomically thin limit.

**Flexoelectric coupling strength from data**

*Effective Young's modulus of a plate with strain gradient elasticity under bending*: There is no general fixed relationship between the locally defined fields $\varepsilon$ and $\nabla\varepsilon$; this relationship depends on the geometry of the object and the geometry of the deformation. However, for the specific case of a plate of thickness $t$ under bending, an effective $E$ may be derived as follows. The Gibbs free energy $G$ of a homogeneous plate comprised of a linearly elastic material is derived from only the strain part of equation (1) of the main text:

$$G = \int \phi dV = \frac{1}{2} E \iint \varepsilon^2 dz dA \tag{7}$$

where $dV$ is a volume element, $dz$ is a line element along the thickness axis, $dA$ is a planar area element, and integrals are over the entire object. For a material with homogeneous strain gradient elasticity, the corresponding expression derived from equation (1) of the main text is:

$$G = \frac{1}{2} E \iint \varepsilon^2 dz dA + \frac{1}{2} K \iint (\nabla \varepsilon)^2 dz dA. \tag{8}$$

However, under a pure bending deformation, the following fixed relationship exists between the integrals with respect to $z$ in equation (8):

$$\int (\nabla \varepsilon)^2 dz = \frac{12}{t^2} \int \varepsilon^2 dz. \tag{9}$$

The validity of equation (9) may be readily observed from a geometric argument (Fig. 1b) and also from our numerical simulations of a circular plate under bending (Extended Data Fig. 5f, for applied force $F = 2$ nN). Inserting equation (9) into equation (8), we obtain:

$$G = \frac{1}{2}\left(E + \frac{12K}{t^2}\right) \iint \varepsilon^2 dzdA. \tag{10}$$

Thus, by comparison of equations (10) and (7), it is apparent that strain gradient elasticity of strength $K$ increases effective Young's modulus for bending a plate by $12K/t^2$, similar to a result derived previously[7].

*Flexoelectric coupling strength of SrTiO3 drumheads*: The thermodynamic free energy density for a flexoelectric material is given as[1] (other strain gradient elasticity effects and piezoelectric effects have been ignored):

$$G = \iint \left(\frac{1}{2\chi}P^2 + \frac{1}{2}E\varepsilon^2 - \frac{\mu}{2\chi}(P\nabla\varepsilon - \varepsilon\nabla P) - PQ - \varepsilon\sigma\right) dzdA \tag{11}$$

where $\chi$ is dielectric permittivity, $Q$ is the electric field, $P$ the polarization and $\sigma$ is the stress. Using Euler-Lagrange minimization with respect to the polarization we obtain:

$$P = \chi Q + \mu\nabla\varepsilon. \tag{12}$$

For flexoelectric strain gradient coupling, we identify $K = \mu^2/\chi$. This expression follows from equating the strain gradient elasticity cost $K(\nabla\varepsilon)^2/2$ to the electrostatic cost $P^2/2\chi$ to polarize the sample to a polarization of $P$ (by substituting equation (12) into equation (11) and neglecting any field terms as there is no external field driving the polarization). Using the room temperature bulk value of $\chi \sim 300\epsilon_0$ for SrTiO3, and attributing the mismatch of $E_{\text{lin}}-E_{\text{cub}} = 156$

GPa ± 49 GPa (at $t$ = 13.7 nm, Fig. 3 of main text) to excess effective Young's modulus $12K/t^2 = 12\mu^2/(\chi t^2)$ (equation (10)), we arrive at the figure of $K$ = 2.2 ± 0.7 µN and $\mu$ ~ 86 ± 13 nCm$^{-1}$ that is quoted in the main text.

**Methods References**

37. Pratt, J. R., Shaw, G. A., Kumanchik, L. & Burnham, N. A. Quantitative assessment of sample stiffness and sliding friction from force curves in atomic force microscopy. *J. Appl. Phys.* **107**, 044305 (2010).

38. Cook, S. M. *et al.* Practical implementation of dynamic methods for measuring atomic force microscope cantilever spring constants. *Nanotechnology* **17**, 2135–2145 (2006).

39. Dhondt, G. CalculiX CrunchiX User's manual. Version 2.6.1 (2013).

40. The FEniCS Project Version 1.5. *Arch. Numer. Softw.* **3**, (2015).

41. Poot, M. Mechanical systems at the nanoscale. *Casimir PhD series*, Technische Universiteit Delft, Delft-Leiden (2009).

42. Vella, D. & Davidovitch, B. Indentation metrology of clamped, ultra-thin elastic sheets. *Soft Matter* **13**, 2264–2278 (2017).

43. Sadeghian, H. *et al.* Characterizing size-dependent effective elastic modulus of silicon nanocantilevers using electrostatic pull-in instability. *Appl. Phys. Lett.* **94**, 221903 (2009).

**Acknowledgments:** We thank Steve Kivelson, Peter Littlewood, Neil Mathur, and Pavlo Zubko for discussions. We also thank Prastuti Singh and Ruijuan Xu for discussions on the polymer-assisted transfer of membranes, and for critically reviewing the manuscript. This work was supported by the U.S. Department of Energy, Office of Basic Energy Sciences, Division of Materials Sciences and Engineering, under contract no. DE-AC02-76SF00515 (synthesis and membrane devices); the Air Force Office of Scientific Research (AFOSR) Hybrid Materials MURI under award no. FA9550-18-1-0480 (elasticity measurements, modeling, and analysis); and the Gordon and Betty Moore Foundation's Emergent Phenomena in Quantum Systems Initiative through grant no. GBMF4415 (initial development of 2D compliance mapping and synthesis equipment). Part of this work was performed at the Stanford Nano Shared Facilities (SNSF), supported by the National Science Foundation under award ECCS-1542152.

**Author contributions:** Sample synthesis and final measurements were performed by V.H. Modeling and analysis were developed and performed by S.C. and V.H. S.S.H., D.L., and V.H. developed the transfer process for making high quality freestanding drumheads, and synthesized initial samples. Y.A.B. developed the 2D linear compliance mapping protocol with S.C. All work was supervised by Y.H. and H.Y.H. V.H., S.C., and H.Y.H. wrote the manuscript with input from all authors.

**Competing interests:** Authors declare no competing interests.

**Data and materials availability:** The codes for the finite element analysis of the bending equation and the 3D FES, and the data that support this study are available upon reasonable request. The summary table of all measured drumheads, their sizes and the Young's modulus and tension is available in the Extended Data.

**Figures and Tables**

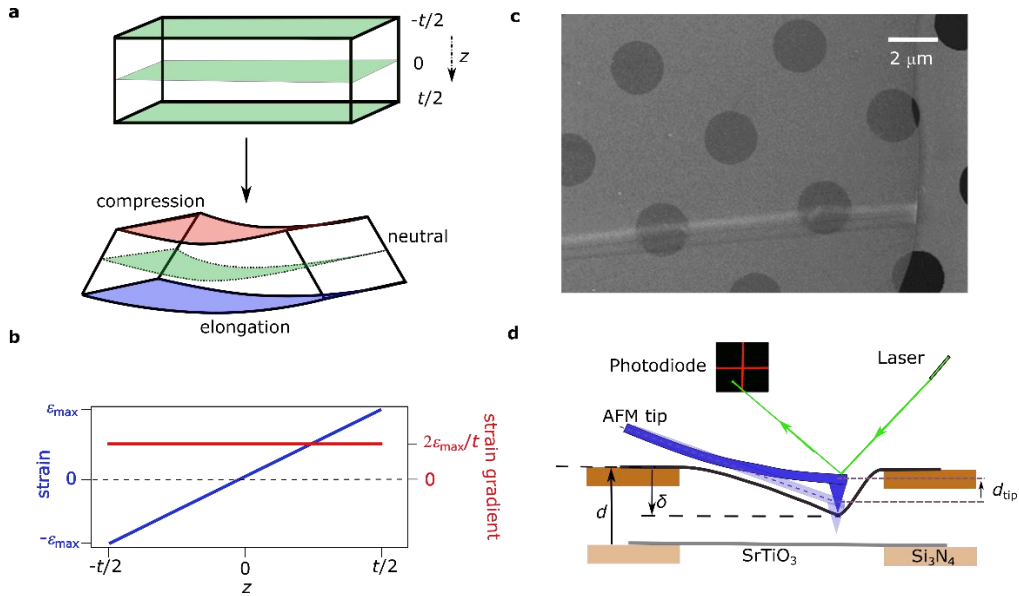

**Fig. 1. Schematic of a bending deformation and experimental geometry. a,** Bending deformation of a plate element of thickness *t*. One side is elongated to a positive strain of $\varepsilon_{max}$, the other side is compressed to a negative strain of $-\varepsilon_{max}$, with a neutral middle plane. **b,** The resulting strain profile in the out-of-plane direction is linear between $\pm\varepsilon_{max}$, while the (uniform) strain gradient is proportional to $\varepsilon_{max}$. The ratio of the mean quadratures of both quantities is independent of $\varepsilon_{max}$, varying in a fixed ratio of $12/t^2$. **c,** Scanning electron micrograph of a 20 nm thick SrTiO$_3$ membrane transferred onto a Si$_3$N$_4$ membrane with an array of 2 μm diameter holes. **d,** Measurement schematic of the drumhead deformation. The faded color elements denote the drumhead and the undeflected AFM tip before tip-sample interaction. As the sample is raised by a distance *d*, the force $F(d)$ causes the tip to deflect by $d_{tip}$ and the membrane under the tip by $\delta$.

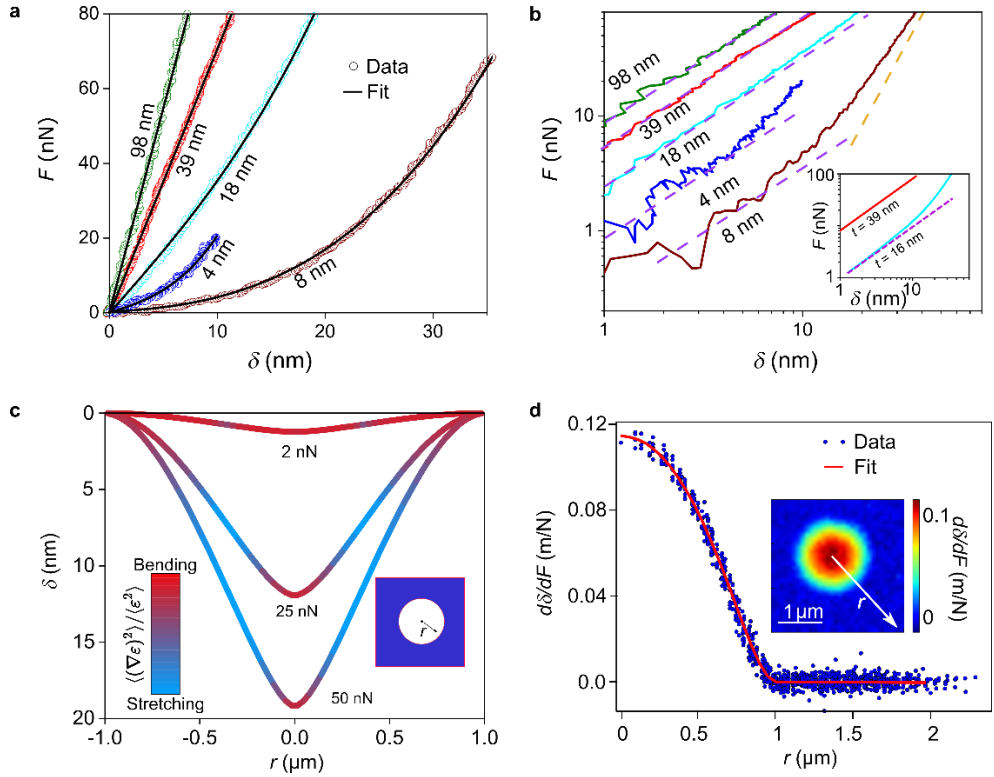

**Fig. 2. Deformation response of freestanding SrTiO$_3$ membranes, experiment and simulation. a,** Representative force-deflection ($F(\delta)$) curves under center-point loading, showing linear and nonlinear elastic regimes depending on SrTiO$_3$ drumhead thickness and magnitude of forcing. For all drumheads $R = 1$ μm, except for $t = 4$ nm ($R = 0.25$ μm) and $t = 98$ nm ($R = 5$ μm). The use of different drumhead radii allows for varying the onset of nonlinearity. Solid lines are fits to the raw data with a cubic polynomial, parameterizing the nonlinearity in $F(\delta)$ for lower thicknesses. **b,** $F(\delta)$ curves re-plotted on a log-log scale showing the crossover between linear and cubic response. Inset shows $F(\delta)$ results of a three-dimensional finite element simulation (3D FES). **c,** 3D FES-calculated drumhead deformation under center-point loading, showing in color scale the relative predominance of bending over stretching, as parameterized by $\langle(\nabla\varepsilon)^2\rangle/\langle\varepsilon^2\rangle$ where the average is taken over the thickness (Methods). Inset shows the drumhead

schematic, defining $r$. **d,** In the linear regime (bending dominated), experimental drumhead compliance $\frac{d\delta}{dF}\big|_{F=0}$ is plotted as a function of $r$ from drumhead center. Fitting to the numerical solution of the biharmonic equation yields pre-tension $T$ and $E_{\text{lin}}$. The inset shows this compliance data as a 2D color map.

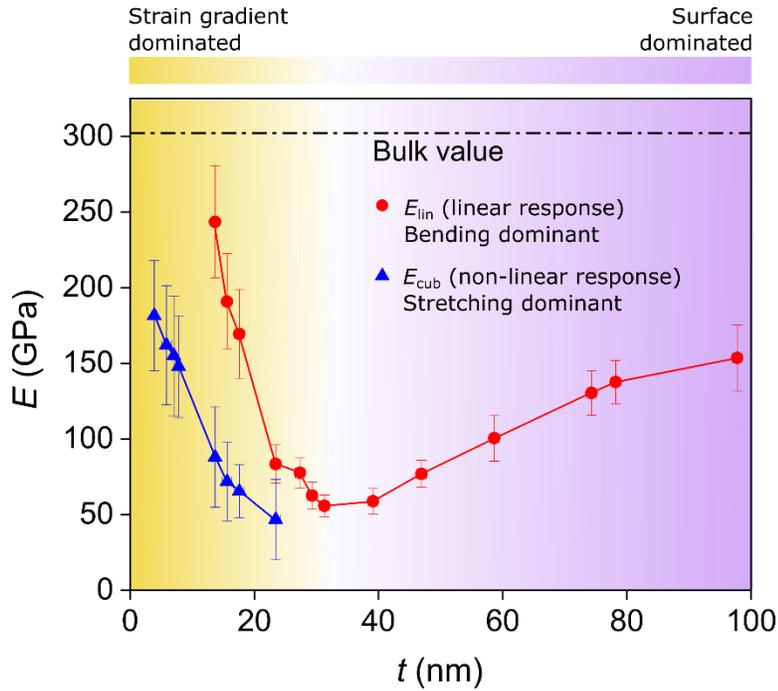

**Fig. 3. Thickness dependence of experimentally extracted effective Young's modulus, for linear and cubic response of the SrTiO₃ membranes.** $E_{\text{lin}}$ and $E_{\text{cub}}$ are Young's modulus inferred from the linear and cubic part of the $F(\delta)$ curves, respectively. The non-monotonic thickness dependence of $E_{\text{lin}}$ arises from dominance of surface elasticity for $t > 31$ nm, and strain gradient elasticity for $t < 31$ nm. $E_{\text{cub}}$ is lower than $E_{\text{lin}}$ for thicknesses where they can both be independently measured, demonstrating higher intrinsic stiffness for bending as compared to stretching. Data from a total of 168 drumheads are represented in this plot. Error bars include an estimated 10% uncertainty in spring constant calibration, and the standard error of the mean. Error bars in $E_{\text{cub}}$ include an additional estimated 12% uncertainty from the 'lookup table' method.

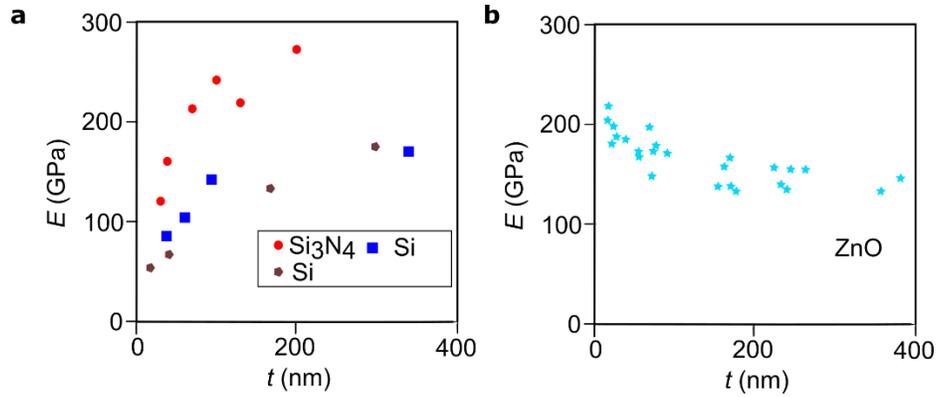

**Extended Data Fig. 1. Compilation of size effects in Young's modulus from literature[10–12,43].** **a,** Materials in the literature studied for size dependence of Young's modulus $E$ showing a monotonic Young's modulus variation as a function of the thickness $t$. All materials in this panel display a deleterious surface effect resulting in a decrease in $E$ as $t$ is decreased. **b,** ZnO shows a monotonic size dependence as well, albeit with a positive surface effect as opposed to those materials shown in **a** resulting in an increase in $E$ as $t$ is decreased.

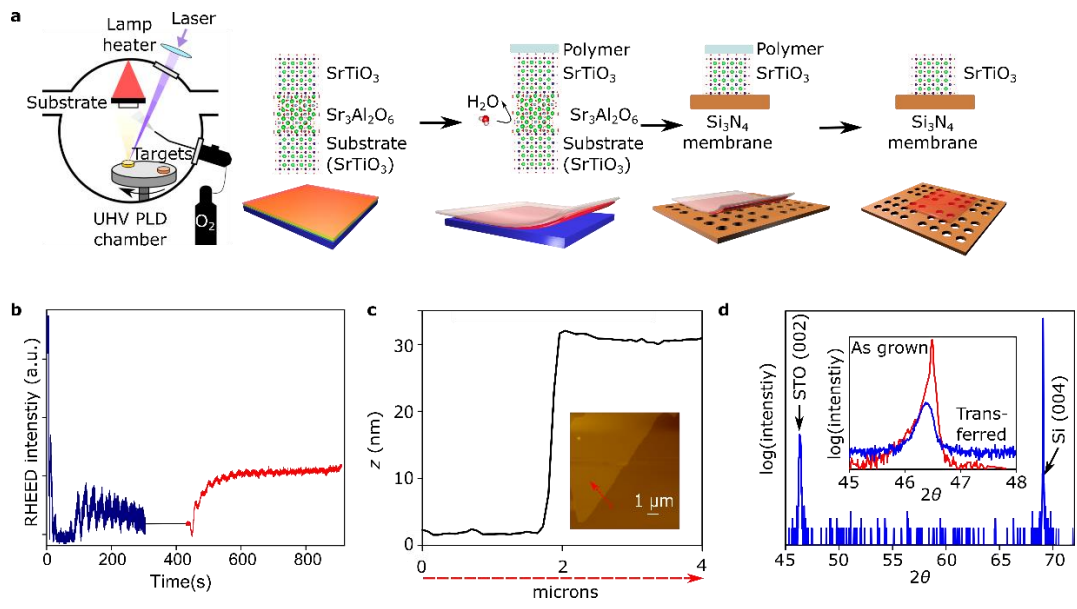

**Extended Data Fig. 2. Fabrication and characterization of SrTiO$_3$ membrane samples. a,** Fabrication of freestanding SrTiO$_3$ drumheads shown in steps (from left to right): pulsed laser deposition; release of the SrTiO$_3$ film from the growth substrate by water etch of the sacrificial Sr$_3$Al$_2$O$_6$ layer; subsequent lift off and transfer onto a commercial Si$_3$N$_4$ membrane with an array of circular holes. The bulk Si supporting frame is not shown here. The last step is removal of the polymer support to produce suspended drumheads. **b,** Reflection high-energy electron diffraction (RHEED) intensity oscillations during Sr$_3$Al$_2$O$_6$ (blue) and SrTiO$_3$ (red) film growth. **c,** Atomic force microscopy (AFM) topography of the transferred membrane. Inset shows topography near a cleaved edge of a 29 nm thick sample, with a line cut along the red arrow revealing the SrTiO$_3$ film thickness. **d,** X-ray diffraction (XRD) $\theta$-$2\theta$ symmetric scan of the transferred membrane of SrTiO$_3$ showing the SrTiO$_3$ (002) peak from the film and the Si (004) peak from the grid support. Inset shows XRD $\theta$-$2\theta$ symmetric scans of the as-grown film and transferred membrane around the SrTiO$_3$ (002) peak. This film is 58.5 nm thick.

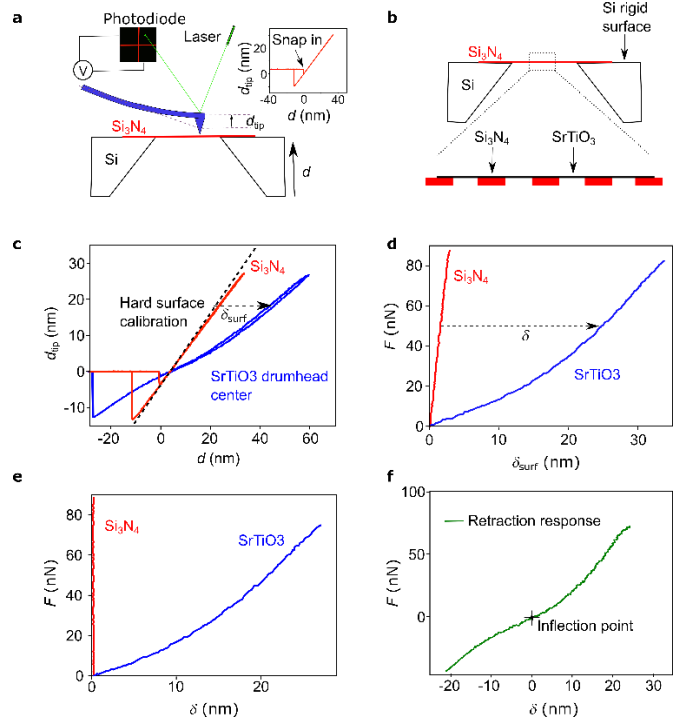

**Extended Data Fig. 3. Measurement procedure and raw data processing. a,** Illustration of raising and impacting the sample surface into the AFM tip. The AFM stage moves a distance $d$ and deflects the tip a distance $d_{\text{tip}}$. Inset shows an example of $d_{\text{tip}}(d)$ on loading and unloading, showing 'snap-in'. **b,** Illustration of the three relevant regions of the sample surface. **c,** Example data for the regions indicated in **b**. The Si rigid surface is used to fix $d_{\text{tip}} = d$ during contact and thus calibrate $d_{\text{tip}}$ (measured via photodiode). Sample surface deflection $\delta_{\text{surf}}$ is then readily found for compliant regions ($Si_3N_4$ and $SrTiO_3$). **d,** Force $F$ as a function of loading-unloading averaged $\delta_{\text{surf}}$. Force is proportional to $d_{\text{tip}}$, such that the vertical axes of **c** and **d** differ only by a constant factor. The $Si_3N_4$ membrane deflection is determined locally around each $SrTiO_3$ drumhead to infer **e**, the deflection $\delta$ of the $SrTiO_3$ alone, by subtraction. **f,** Identification of the zero-deflection point of the $SrTiO_3$ drumhead, via the inflection point of the retract curve only.

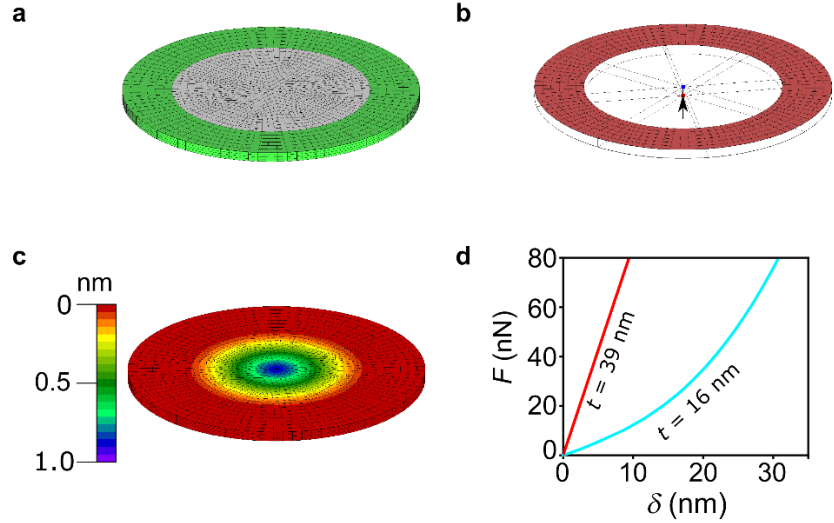

**Extended Data Fig. 4. Finite element modeling of freestanding drumheads. a,** Finite element mesh. The grey region represents the freestanding SrTiO$_3$ drumhead, and the green surrounding region represents the portion of the SrTiO$_3$ membrane that is directly supported by Si$_3$N$_4$. **b,** The red shaded area represents the region of the SrTiO$_3$ membrane surface that is in direct contact with Si$_3$N$_4$; clamped boundary conditions were imposed in this region. The red node experiences a force $F$ in the upward direction, representing the impacting AFM tip. The corresponding displacement of the blue node represents the under-tip membrane deflection $\delta$. **c,** A color map representing all nodal displacements $u(r)$ in response to the applied force at the central node. **d,** The simulated $F(\delta)$ captures linear and nonlinear regimes, depending on the applied force and the thickness $t$ of the SrTiO$_3$ drumhead. This simulation is for a Hookean material in the elastic regime; i.e. $E_{lin}$ is the same as $E_{cub}$ which is the same as Young's modulus $E$, taken here to be 50 GPa for the 16 nm thick film and 40 GPa for the 39 nm thin film.

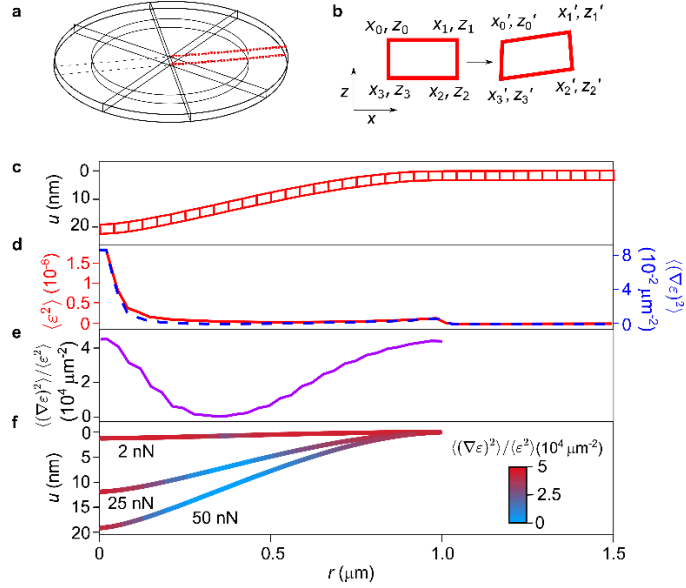

**Extended Data Fig. 5. Analysis of the nature of the deformation in a SrTiO$_3$ drumhead. a,** The displacements of the indicated nodes of the finite element model were analyzed. **b,** Each set of four adjacent nodes defined a rectangular 'cell' of undeformed corner coordinates ($x_i$, $z_i$) and deformed corner coordinates ($x_i'$, $z_i'$) [$0 \leq i \leq 3$]. **c,** The complete set of 'cells' defined by the nodes from **a** for a drumhead of thickness of 15.6 nm, Young's modulus of 100 GPa, and radius 1 μm, experiencing an applied force of 50 nN (a regime corresponding to nonlinear $F(\delta)$). Cell deflections $u$ are represented as a function of radial coordinate $r$. Vertical cell dimensions are scaled for clarity. **d,** For each 'cell' in **c** the average squared strain $\langle \varepsilon^2 \rangle$ and average squared strain gradient $\langle (\nabla \varepsilon)^2 \rangle$ across the thickness of the cell were calculated. These properties are plotted as a function of radial coordinate $r$ (at the cell center). **e,** Ratio of $\langle (\nabla \varepsilon)^2 \rangle / \langle \varepsilon^2 \rangle$. Areas where $\langle \varepsilon^2 \rangle$ and $\langle (\nabla \varepsilon)^2 \rangle$ are in maximum proportion correspond to bending. Otherwise, a degree of stretching is implied. Thus, for the present simulation, stretching is concentrated at intermediate $r$. **f,** Drumhead deflection $u(r)$ for $F = 2$, 25 and 50 nN which is the same as the main text Fig. 2C. The color scale for $F = 50$ nN is the data from **e**.

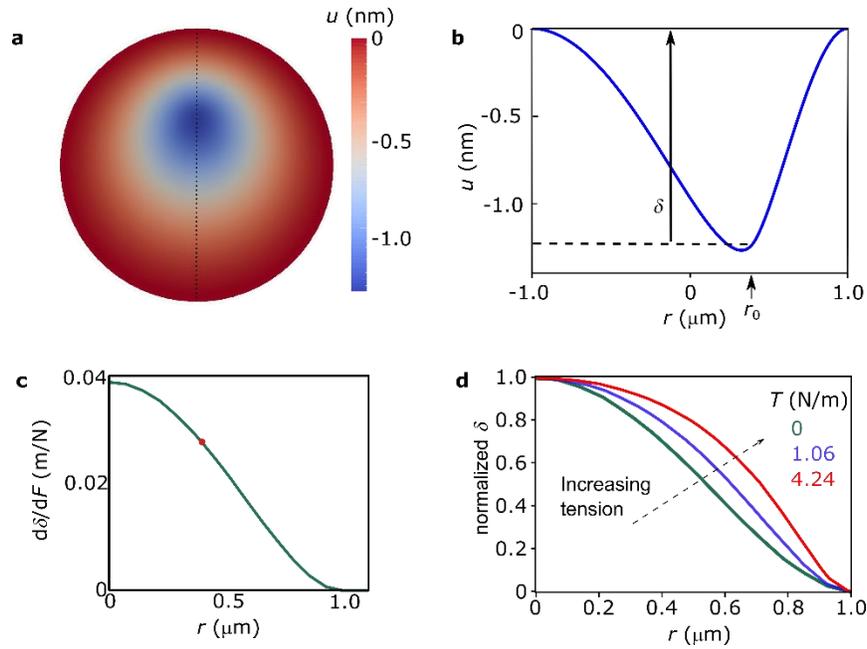

**Extended Data Fig. 6. Numerical solutions of the bending equation. a,** Color map showing simulated deformation $u(r)$ of a 1 μm radius, 40 nm thick membrane drumhead loaded by 50 nN at a lateral position $r_0 = 0.4$ μm (with respect to the center). For this example, Young's modulus $E = 100$ GPa and pre-tension $T$ is absent. **b,** Diametric line cut of the data in **a** along the dotted line shown in **a**. Simulated under-tip deflection $\delta$ is indicated. **c,** Force-normalized under-tip deflection $d\delta/dF$ (i.e. membrane compliance) on varying loading position. Thus, multiple simulated datasets similar to those in **a** and **b** were calculated to construct the plot shown in **c**. The red dot indicates the datapoint derived from the simulation in **a** and **b**. **d,** Functional form of normalized compliance curves as a function of increasing tension. Note that the green, blue and red lines represent values of $\lambda = 0$, 4 and 8 respectively, where $\lambda = \sqrt{\frac{TR^2}{D}}$ is a convenient and universal parameterization of the dominance of tension over bending rigidity $D$[41]. Here representative $T$ for $t = 20$ nm, $E = 100$ GPa are given; however, these curves are universal, so scaling $T$, $E$ and $t$ appropriately to preserve $\lambda$ does not change the curves.

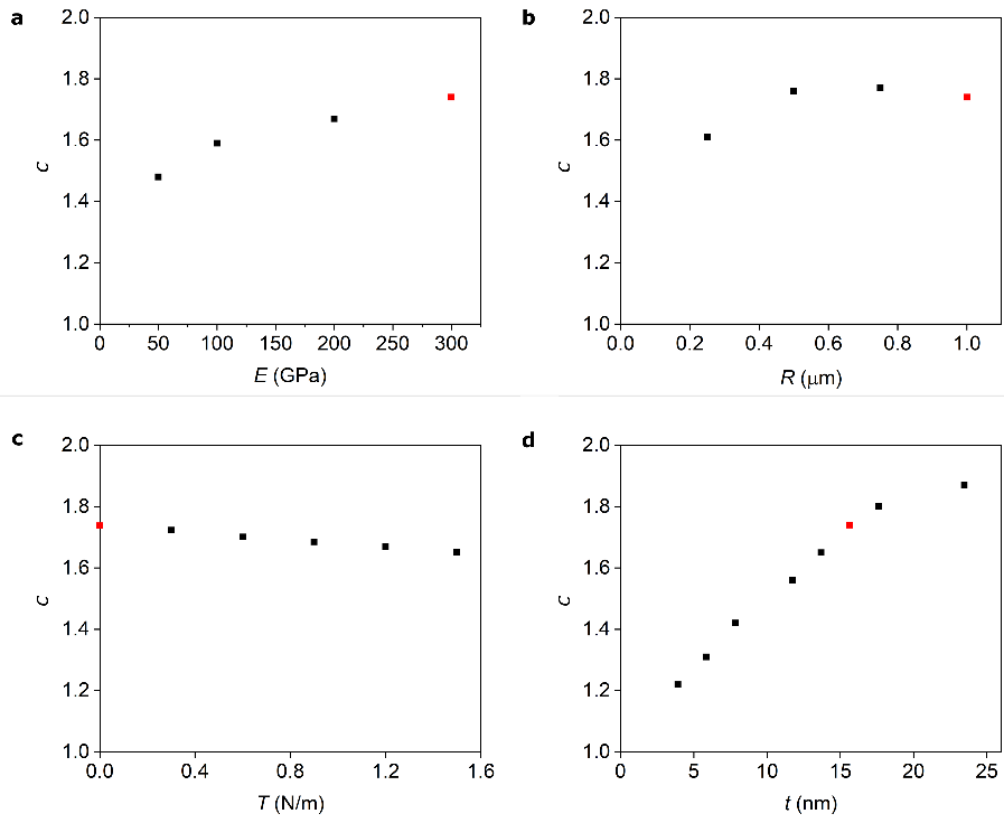

**Extended Data Fig. 7. Nonlinear force-deflection response on changing the relevant experimental parameters. a-d**, The effect on $c$ on varying each parameter $E$, $R$, $T$, $t$ in turn, keeping the other three constant. The variables are $R$, the drumhead radius, $t$, the thickness of the membrane, $E$, is Young's modulus and $T$, the pre-tension. The vertical axes plot $c$, where $c$ is the relevant coefficient obtained by fitting a cubic polynomial to simulated force-deflection data (equation (2) of main text). Red data points show the common points on all plots.

| Thickness (nm) | Radius (μm) | Number of drumheads | $E_{cub}$ (GPa) | $E_{lin}$ (GPa) | $T$ (N/m) |
|---|---|---|---|---|---|
| 4 | 0.25 | 13 | 182 ± 36 | - | - |
| 6 | 0.25 | 7 | 162 ± 39 | - | - |
| 7 | 0.25 | 6 | 155 ± 39 | - | - |
| 7.8 | 0.25 | 4 | 152 ± 37 | - | - |
| 7.8 | 1 | 9 | 147 ± 34 | - | - |
| 13.7 | 1 | 8 | 88 ± 33 | 244 ± 37 | 0.4 ± 0.2 |
| 15.6 | 1 | 6 | 72 ± 26 | 190 ± 32 | 0.7 ± 0.1 |
| 17.6 | 1 | 7 | 66 ± 18 | 169 ± 29 | 0.3 ± 0.2 |
| 23.4 | 1 | 7 | 47 ± 27 | 84 ± 14 | 0.1 ± 0.1 |
| 27.3 | 1 | 11 | - | 77 ± 10 | 0.5 ± 0.3 |
| 29.3 | 1 | 14 | - | 63 ± 9 | 1.6 ± 1.1 |
| 31.2 | 1 | 13 | - | 56 ± 7 | 1.1 ± 0.6 |
| 39.1 | 1 | 13 | - | 59 ± 8 | 0.3 ± 0.4 |
| 46.8 | 1 | 9 | - | 77 ± 9 | 0.9 ± 0.5 |
| 46.8 | 5 | 4 | - | 76 ± 8 | 0.7 ± 0.7 |
| 58.5 | 1 | 11 | - | 101 ± 16 | 6.5 ± 1.8 |
| 74.3 | 5 | 11 | - | 131 ± 15 | 4.2 ± 0.8 |
| 78.2 | 5 | 7 | - | 138 ± 15 | 1.2 ± 2.0 |
| 97.6 | 5 | 8 | - | 154 ± 22 | 0.4 ± 0.9 |

**Table S1. Table of measurements and evaluated values of Young's modulus and pre-tension.** This table summarizes all of the data taken on 168 $SrTiO_3$ drumheads of different $R$ and $t$. The error bars in $E_{lin}$ subsumes error from AFM tip calibration and statistical error around the mean. $E_{cub}$ further has a 12% error included from the 'lookup table' method. The pre-tensions quoted here correspond to ~ 0.1% pre-strain at maximum, which can arise during the transfer process. The error bars for the pre-tension represent the range of pre-tensions observed in each set of experiments and are the statistical standard deviation around the mean.

Supplementary Information for

# Giant strain gradient elasticity in SrTiO$_3$ membranes: bending versus stretching

V. Harbola*[‡1,2], S. Crossley*[‡2,3], S. S. Hong[2,3], D. Lu[1,2], Y. A. Birkhölzer[4], Y. Hikita[2] & H. Y. Hwang*[2,3]

[1]*Department of Physics, Stanford University, Stanford, California 94305, USA*
[2]*Stanford Institute for Materials and Energy Sciences, SLAC National Accelerator Laboratory, 2575 Sand Hill Road, Menlo Park, California 94025, USA*
[3]*Department of Applied Physics, Stanford University, Stanford, California 94305, USA*
[4]*Department of Inorganic Materials Science, Faculty of Science and Technology and MESA+ Institute for Nanotechnology, University of Twente, P.O. Box 217, 7500 AE Enschede, The Netherlands*

[‡]These authors contributed equally: V. Harbola and S. Crossley
Email: varunh@stanford.edu, samuelcrossley@gmail.com, hyhwang@stanford.edu

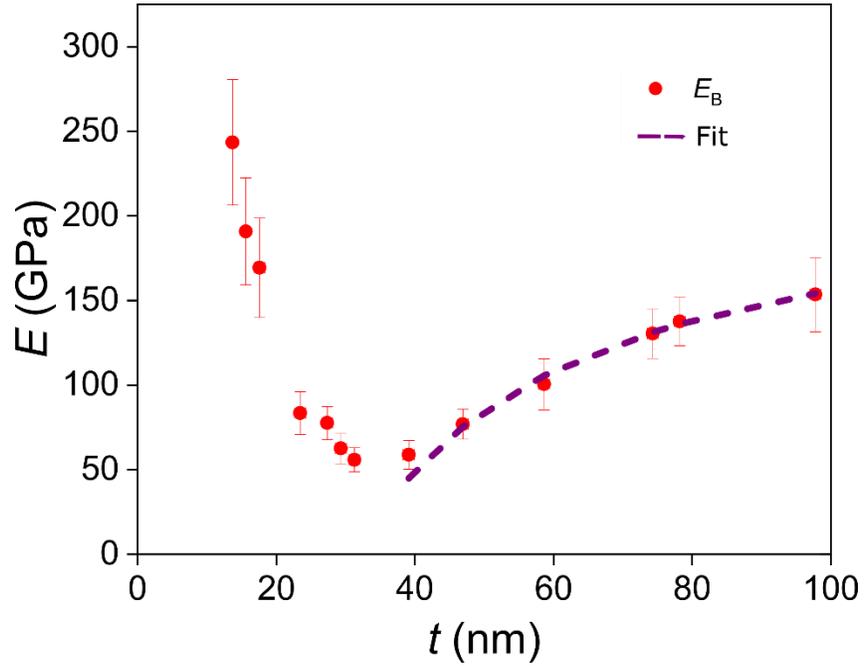

**Supplementary Figures**

**Supplementary Fig. 1. Surface elasticity in thicker SrTiO₃ membranes.** The trend of the elastic modulus in thicker membranes of $SrTiO_3$ (up to our measurement limit of 98 nm) can be well described by the surface elasticity contribution as discussed in the main text and Supplementary Data Fig. 1. The fit function is of the form *a+b/t* to account for surface elasticity and describes the experimental data well for membranes with $t > 40$ nm.

**Supplementary Notes**

**Review of size effects in elasticity**

Li and coworkers[1] employed patterning and under-etch methods to make cantilevers of Si, with lateral dimensions of order 10s of microns. Effective Young's modulus under bending, deduced by vibration spectroscopy, was observed to monotonically increase from ~50 GPa to ~170 GPa as thickness was increased from 12 nm to 300 nm (Extended Data Fig. 1a). A mechanism of surface elasticity was proposed.

Sadeghian and coworkers[2] employed patterning and under-etch methods to make cantilevers of Si, with lateral dimensions of order 10s of microns. Effective Young's modulus under bending, deduced by slow electrostatic loading, was observed to monotonically increase from ~85 GPa to ~170 GPa as thickness was increased from ~50 nm to 1000 nm (Extended Data Fig. 1a). A mechanism of surface elasticity was proposed.

Gavan and coworkers[3] employed patterning and under-etch methods to make cantilevers of $Si_3N_4$, with lateral dimensions of order 10s of microns. Effective Young's modulus under bending, deduced by vibration spectroscopy, was observed to monotonically increase from ~140 GPa to ~290 GPa as thickness was increased from 20 nm to 684 nm (Extended Data Fig. 1b). A mechanism of surface elasticity was proposed. Furthermore, surface stress was ruled out by the data.

Chen and coworkers[4] employed a controlled thermal evaporation procedure to make nanowires of [0001]-oriented ZnO with lateral dimensions of order microns. Effective Young's modulus under bending, deduced by vibration spectroscopy, was observed to monotonically decrease from ~220 GPa to ~140 GPa as diameter was increased from 17 nm to 550 nm (Extended Data Fig. 1b). A mechanism of surface elasticity was proposed.

For few-layer graphene, through studying the drumhead response using AFM, Poot and coworkers established Young's modulus of 920 GPa for graphene, claiming a thickness independence over a thickness range of 8-100 layers of graphene[5].

For $MoS_2$, Castellanos-Gomez and colleagues[6] measured Young's modulus through the nonlinear force-deflection response and found it to be 330 GPa over a thickness range of 5-25 layers of $MoS_2$. However they do raise the question about size dependent Young's modulus in that material as Young's modulus for bulk $MoS_2$ is 240 GPa[7].

In summary, large monotonic size effects of elasticity have been reported in several materials outside the family of complex oxides (Extended Data Fig. 1). The most convenient fabrication approaches typically exploit semiconductor and micro-electro-mechanical device (MEMS) processes, such that archetypal semiconductor and MEMS materials are those most systematically studied thus far. Surface elasticity is the widely accepted mechanism of the monotonic size effects reported. The effects of surface elasticity are amplified by bending deformation, ubiquitous in cantilever studies, given that bending is disproportionately resisted by surface regions that are maximally displaced from the neutral axis (Fig. 1a, b of the main text).

**Surface dominated elasticity effects in thicker $SrTiO_3$ films**

Surface effects in elasticity can arise both from a surface stress (strain independent surface stress) which has been show to exist in freestanding perovskite lamellas[8], and a different elastic modulus at the surface (strain dependent surface stress) of the material compared to the bulk, known as surface elasticity[3]. As discussed in the previous section, surface elasticity has been able to describe thickness dependent trends in materials, and surface stress has been shown to have limited contributions[3]. The role of surface elasticity in most materials is to reduce the elastic modulus as the thickness is decreased (Extended Data Fig. 1a). We observe a similar trend in SrTiO$_3$ for thicknesses > 40 nm but the elastic modulus observed for the 98 nm film was about half the value of that observed in bulk SrTiO$_3$. In bulk, SrTiO$_3$ undergoes a cubic→tetragonal (Pm$\bar{3}$m→I4/mcm) phase transition below 105 K. This transition is only improperly ferroelastic, limiting the transitional elastic anomaly to low 10s of percent once domain wall mobility is accounted for[9,10]. The ~200% change in $E_{lin}$ we observe for $t > 31$ nm is larger than this, however the possibility that the surface elasticity is due to Pm$\bar{3}$m→I4/mcm may be considered. We can fit the data in this thicker regime with a function $a+b/t$ where $a$ represents the asymptotic value of the Young's modulus in the bulk limit and $b$ is representative of the difference in elasticity between the surface and the bulk (surface elasticity). This fit is shown in Supplementary Fig. 1. The asymptotic value $a$ from the fit is 228 GPa, which is roughly 80% of the bulk value of SrTiO$_3$ (ref. [9]). A direct measurement of thicker membranes is limited by the increasing contribution of the deflection of the Si$_3$N$_4$ support.

**Relationship of $E_{cub}$ to effective Young's modulus for pure collinear stretching**

Effective Young's Modulus $E_{cub}$ is defined as the modulus that a linearly elastic, homogeneous drumhead would need to have in order to produce the observed value of the fitted nonlinear force-deflection ($F(\delta)$) response. Although nonlinear $F(\delta)$ response occurs due to stretching, simulation confirms that bending is still present in the overall deformation geometry (Extended Data Fig. 5f), and $E_{\text{cub}}$ is not formally equivalent to Young's modulus that would be found from a pure collinear stretching experiment.

To understand that $E_{\text{cub}}$ is an upper bound for the collinear stretching modulus for a drumhead including strain gradient elasticity, one can consider how the deformation of a drumhead in the absence of strain gradient elasticity (Extended Data Fig. 5f) would be modified upon activating a perturbative strain gradient elasticity.

In the linearly elastically deformed shape, bending and associated strain gradients are largest close to the center of the drumhead, and close to the edges (Fig. 2c and Extended Data Fig. 5f). However, the edge regions encompass a far larger volume of material. If we could turn strain gradient elasticity on perturbatively, global energy minimization will cause the system to act to reduce the volume of material that experiences a strain gradient. Due to the linear radial dependence of the circumference, this can be achieved by shifting the bending at the edges of the drumhead closer in towards the center of the drumhead. The effective drumhead radius would therefore be reduced on activation of strain gradient elasticity, requiring a smaller value of $E_{\text{cub}}$ to explain a given experimental value of the coefficient to the cubic term in equation (2) of the main text. Therefore, the quoted $E_{\text{cub}}$ (based on an overstated effective radius $R$) represents an upper bound on the intrinsic stretching modulus, with a mismatch that scales with the strength of strain gradient elasticity that is present.

**Flexoelectricity and SrTiO$_3$ elasticity**

Ever since experiments on flexoelectricity have been performed, there has been a large discrepancy between the phenomenological estimates for the flexocoupling coefficient[11,12] and what has been observed experimentally[13–16]. Bulk SrTiO$_3$ was one of the only materials in these studies to show a flexocoupling coefficient of $|f| = 2.6$ V lying within the Kogan limit of 1-10

V[17]. However recent work in SrTiO$_3$ thin films measuring the flexoelectric response through a tunneling measurement[18] demonstrates an enhanced flexocoupling coefficient of ~25 V, similar to what we obtain from our elastic measurements, assuming that complete strain gradient elasticity is a result of the flexoelectric response of the SrTiO$_3$ membrane. There is no broadly accepted explanation for flexocoupling in excess of the Kogan limit. One possibility at the nanoscale is that surface piezoelectricity acts in opposition to bulk flexoelectricity and comes to dominate as the latter effect drops to zero at the nanoscale, but the two act to cancel each other out in the bulk[19]. Another possibility in general is the electronic contribution to the flexoelectric effect, although simulations fail to account for the required magnitude of enhancement[20]. For SrTiO$_3$, another way this could manifest itself is that surface piezoelectricity itself has some size dependence related to the emergent polarity in thin SrTiO$_3$[21]. Recent work[18] presents some evidence for higher-than-linear order flexocoupling activated by extremely large strain gradients in SrTiO$_3$. However, we note that the largest strain gradients of ~$3\times10^5$ m$^{-1}$ applied in the present work are two orders-of-magnitude smaller and are yet associated with a similar nominal flexocoupling coefficient. We also note that we use the value of bulk permittivity ($300\epsilon_0$) for SrTiO$_3$ to obtain the value of the flexoelectric coefficient from the strain gradient coupling. This value can be different for thin films (generally smaller) and may be influenced by polar nano-regions that have been reported to occur in some cases[21]. However, $|f|$ has a relatively weak dependence on the permittivity ($|f| \sim \chi^{-1/2}$), such that a very large value of ~$40000\epsilon_0$ would be required to evaluate a bulk value of $|f| = 2.6$ V from our elasticity data.

**Glossary of variables**

*Description of drumhead*

$t$ : Thickness of SrTiO$_3$ drumhead.

$R$ : Radius of SrTiO$_3$ drumhead.

$r$ : Radial coordinate with respect to the center of the drumhead.

$u(r)$ : Deflection profile of drumhead from its neutral plane.

*Force microscopy*

$F$ : Total force applied to the SrTiO$_3$ drumhead by the atomic force microscope (AFM) tip.

$p$ : Pressure applied to the SrTiO$_3$ drumhead by the AFM tip. Integral of $p$ over all space is equal to $F$.

$r_0$ : Radial coordinate of atomic force microscope tip, with respect to the center of the drumhead.

$\delta$ : Under-tip deflection of the SrTiO$_3$ drumhead from its neutral plane. Equivalent to $u(r_0)$.

$\delta_{\text{sur}}$ : Under-tip deflection of the sample surface.

$d_{\text{tip}}$ : Deflection of the AFM tip.

$d$ : Distance moved by sample, with respect to an arbitrary origin.

$k$ : Force constant of the AFM tip.

*Elasticity*

$E$ : Young's modulus. The 'local' $E$ is a field-like quantity that might vary across regions of an inhomogeneously elastic object. The 'effective' $E$ is the result of a particular experiment (such as bending, or stretching) on an object as a whole, where experimental data are interpreted *as if* that object were comprised of a linearly and homogeneously elastic material of Young's modulus $E$.

$E_{\text{cub}}$ : Effective Young's modulus inferred from a stretching-like response whose experimental signature is nonlinear $F(\delta)$. Represents an upper bound on effective Young's modulus for collinear stretching.

$E_{\text{lin}}$ : Effective Young's modulus inferred from a pure bending deformation whose experimental signature is linear $F(\delta)$. Nominally, for a material without a strain gradient elastic coupling, $E_{\text{lin}} = E_{\text{cub}}$.

$\nu$ : Poisson's ratio of the SrTiO$_3$ drumhead, assumed equal to 0.23. Reasonable variations in $\nu$ do not significantly impact the evaluated $E_{\text{cub}}$ or $E_{\text{lin}}$.

$T$ : Pre-tension present in the SrTiO$_3$ drumhead.

$D$ : Bending rigidity of the SrTiO$_3$ drumhead (depends on drumhead geometry, $\nu$ and $E_{\text{lin}}$).

$K$ : Strain gradient elasticity coupling coefficient.

$\varepsilon$ : Strain along the radial in-plane direction of the drumhead.

$(\nabla\varepsilon)$ : 'Strain gradient': derivative of strain along the radial in-plane direction with respect to the vertical out-of-plane direction.

*Flexoelectricity*

$\mu$ : Effective flexoelectric coupling coefficient, defined as the electric polarization induced by unit strain gradient.

$\chi$ : Absolute permittivity of a material.

$\epsilon_0$ : Permittivity of free space.

$f$ : Flexocoupling coefficient, $\mu/\chi$.

$Q$: The electric field.

$\sigma$: The local stress during the deformation of a material.

$P$: Polarization in a material. In this manuscript the polarization discussed is a result of strain gradients arising from flexoelectric coupling.

**Supplementary References**

1. Li, X., Ono, T., Wang, Y. & Esashi, M. Ultrathin single-crystalline-silicon cantilever resonators: Fabrication technology and significant specimen size effect on Young's modulus. *Appl. Phys. Lett.* **83**, 3081–3083 (2003).

2. Sadeghian, H. *et al.* Characterizing size-dependent effective elastic modulus of silicon nanocantilevers using electrostatic pull-in instability. *Appl. Phys. Lett.* **94**, 221903 (2009).

3. Babaei Gavan, K., Westra, H. J. R., Van Der Drift, E. W. J. M., Venstra, W. J. & Van Der Zant, H. S. J. Size-dependent effective Young's modulus of silicon nitride cantilevers. *Appl. Phys. Lett.* **94**, 233108 (2009).

4. Chen, C. Q., Shi, Y., Zhang, Y. S., Zhu, J. & Yan, Y. J. Size dependence of Young's modulus in ZnO nanowires. *Phys. Rev. Lett.* **96**, 075505 (2006).

5. Poot, M. & Van Der Zant, H. S. J. Nanomechanical properties of few-layer graphene membranes. *Appl. Phys. Lett.* **92**, 63111 (2008).

6. Castellanos-Gomez, A. *et al.* Elastic properties of freely suspended MoS$_2$ nanosheets. *Adv. Mater.* **24**, 772–775 (2012).

7. Feldman, J. L. Elastic constants of 2H-MoS$_2$ and 2H-NbSe$_2$ extracted from measured dispersion curves and linear compressibilities. *J. Phys. Chem. Solids* **37**, 1141–1144 (1976).

8. Luk'yanchuk, I. A., Schilling, A., Gregg, J. M., Catalan, G. & Scott, J. F. Origin of ferroelastic domains in free-standing single-crystal ferroelectric films. *Phys. Rev. B* **79**, 144111 (2009).

9. Carpenter, M. A. Elastic anomalies accompanying phase transitions in (Ca,Sr)TiO$_3$ perovskites: Part I. Landau theory and a calibration for SrTiO$_3$. *Am. Mineral.* **92**, 309–327 (2007).

10. Kityk, A. *et al.* Low-frequency superelasticity and nonlinear elastic behavior of crystals. *Phys. Rev. B* **61**, 946 (2000).

11. Kogan, S. M. Piezoelectric effect during inhomogeneous deformation and acoustic scattering of carriers in crystals. *Sov. Phys. Solid State* **5**, 2069–2070 (1964).

12. Tagantsev, A. K. Piezoelectricity and flexoelectricity in crystalline dielectrics. *Phys. Rev. B* **34**, 5883–5889 (1986).

13. Ma, W. & Cross, L. E. Observation of the flexoelectric effect in relaxor Pb(Mg$_{1/3}$Nb$_{2/3}$)O$_3$ ceramics. *Appl. Phys. Lett.* **78**, 2920–2921 (2001).

14. Ma, W. & Cross, L. E. Flexoelectric polarization of barium strontium titanate in the paraelectric state. *Appl. Phys. Lett.* **81**, 3440–3442 (2002).

15. Ma, W. & Cross, L. E. Strain-gradient-induced electric polarization in lead zirconate titanate ceramics. *Appl. Phys. Lett.* **82**, 3293–3295 (2003).

16. Narvaez, J., Vasquez-Sancho, F. & Catalan, G. Enhanced flexoelectric-like response in oxide semiconductors. *Nature* **538**, 219–221 (2016).

17. Zubko, P., Catalan, G. & Tagantsev, A. K. Flexoelectric effect in solids. *Annu. Rev. Mater. Res.* **43**, 387–421 (2013).

18. Das, S. *et al.* Enhanced flexoelectricity at reduced dimensions revealed by mechanically tunable quantum tunnelling. *Nat. Commun.* **10**, 537 (2019).

19. Stengel, M. Surface control of flexoelectricity. *Phys. Rev. B* **90**, 201112 (2014).

20. Hong, J. & Vanderbilt, D. First-principles theory and calculation of flexoelectricity. *Phys. Rev. B* **88**, 174107 (2013).

21. Jang, H. W. *et al.* Ferroelectricity in strain-free $SrTiO_3$ thin films. *Phys. Rev. Lett.* **104**, 197601 (2010).